# Morphotropic Phase Boundary in Sm-Substituted BiFeO$_3$ Ceramics: Local vs Microscopic Approaches

A. Pakalniškis[a], R. Skaudžius[a], D.V. Zhaludkevich[b], A.L. Zhaludkevich[b], D.O. Alikin[c], A.S. Abramov[c], T. Murauskas[a], V.Ya. Shur[c], A.A. Dronov[d], M.V. Silibin[d,e,f], A. Selskis[g], R. Ramanauskas[g], A. Lukowiak[h], W. Strek[h], D.V. Karpinsky[b,j], A. Kareiva[a]

[a]*Institute of Chemistry, Vilnius University, Naugarduko 24, LT-03225 Vilnius, Lithuania*
[b]*Scientific-Practical Materials Research Centre of NAS of Belarus, 220072 Minsk, Belarus*
[c]*School of Natural Sciences and Mathematics, Ural Federal University, Russia*
[d]*National Research University of Electronic Technology "MIET", 124498 Moscow, Russia*
[e]*Institute for Bionic Technologies and Engineering, I.M. Sechenov First Moscow State Medical University, 119991 Moscow, Russia*
[f]*Scientific-Manufacturing Complex "Technological Centre", Zelenograd, 124498 Moscow, Russia*
[g]*Center for Physical Sciences and Technology, LT-02300 Vilnius, Lithuania*
[h]*Institute of Low Temperature and Structure Research, Polish Academy of Sciences, Okolna 2, PL-50422 Wroclaw, Poland*
[j]*South Ural State University, 454080 Chelyabinsk, Russia*

Corresponding author:
A. Pakalniškis (pakalniskis.andrius@chgf.vu.lt);

**Abstract**

Samarium substituted bismuth ferrite (BiFeO$_3$) ceramics prepared by sol-gel synthesis method were studied using both local scale and microscopic measurement techniques in order to clarify an evolution of the crystal structure of the compounds across the morphotropic phase boundary region. X-ray diffraction analysis, transmission and scanning electron microscopies, XPS, EDS/EDX experiments and piezoresponse force microscopy were used to study the structural transitions from the polar active rhombohedral phase to the anti-polar orthorhombic phase and then to the non-polar orthorhombic phase, observed in the Bi$_{1-x}$Sm$_x$FeO$_3$ compounds within the concentration range of 0.08 ≤ x ≤ 0.2. The results obtained by microscopic techniques testify that the compounds in the range of 0.12 ≤ x ≤ 0.15 are characterized by two phase structural state formed by a coexistence of the rhombohedral and the anti-polar orthorhombic phases; two phase structural state observed in the compounds with 0.15 < x < 0.18 is associated with a coexistence of the anti-polar orthorhombic and the non-polar orthorhombic phases. Local scale measurements have revealed a notable difference in the concentration range ascribed to the morphotropic phase boundary estimated by microscopic measurements, the obtained results testify a wider concentration range ascribed to a coexistence of different structural phases, the background of the mentioned difference is discussed.

**Keywords**: sol-gel processing; morphotropic phase boundary; XRD, TEM, SEM, PFM.

1. Introduction

The term morphotropic phase boundary (MPB) usually refers to a concertation region where different crystal structures with a low energy barrier can coexist [1,2]. The most common and well understood MPB exists between the polar rhombohedral (R3m) and tetragonal (P4mm) structures in lead zirconate titanate [3,4]. While such transitions are not allowed by symmetry rules, recently it has been explained that they are mediated by a formation by transitionary monoclinic phase [5]. This phase coexistence region has attracted a lot of attention from scientists as it has been discovered that such compounds show large sensitivity to external stimuli [1,6]. However, while currently dominant piezoelectric materials with MPB (PZT, PMN-BT) are extremely efficient they contain lead which is highly toxic [7]. Due to this drawback and overall shift towards green and sustainable chemistry, which was partly induced by anti Pb legislations, a search for a greener alternative is ongoing [8,9].

The search for a new type of materials led scientists to a discovery of the polar to non-polar transition containing MPB. This significantly improved the approach of new lead free ferroelectric material design with properties similar to PZT [10]. While, in accordance with the Goldschmidt's rule, both polar to polar and polar to non-polar MBPs are fundamentally different [11]. As for the case of polar to polar MPB, the enhancement of the specific properties is related to the easing of polarization rotation, and for polar to non-polar with polarization extension mechanism [10]. Interestingly, it was also estimated that if an intermediate phase exists in the "polar to non-polar" or "polar to polar" MPB, a combination of both mechanisms would be possible. Thus in such case an exponential improvement of the properties could be expected for these materials [8,12,13].

Over the last couple of decades many different compounds were analyzed in search of the aforementioned characteristics. $BiFeO_3$, a lead free multiferroic compound with perovskite type structure, was found to have a polar to non-polar morphotropic phase boundary upon doping with RE ions (La – Lu) [14,15]. The undoped compound is characterized by polar active rhombohedral structure described by R3c space group as well as magnetically active sublattice formed by iron ions which makes it to be room temperature multiferroic with Currie temperature of ~1100 K and Neel temperature of 643 K [2,16,17]. There are some drawbacks specific for the initial compound, viz. bismuth ferrite is characterized by a large leakage current as well as difficulties in preparation of single phase materials [16,18]. Doping of the initial compound with rare earth ions has shown to at least partly solves these problems as well causes a polar (R3c) to non-polar (Pnma) structural transition, with an intermediate $PbZrO_3$-like anti-polar (Pbam) phase [2,19]. While the concentration range of MPB varies depending on the type of RE ion, the stabilization of the anti-polar phase has only been possible with RE ions up to Sm within the MPB which is very narrow among similar oxide systems ~1% [20, 21]. Moreover, correct determination of morphotropic phase boundary is quite difficult because it is strongly dependent on preparation technique and post synthesis treatment [15, 22–24]. It should be noted, that structural characterization of MPB region performed based on microscopic measurements such as XRD, Raman or IR spectroscopy can give different results as compared to local measurements such as TEM or PFM, which poses a great problem for further scientific research [25–27].

Hence the main aim of this work was to investigate the morphotropic phase boundary region in Sm-doped BiFeO$_3$ compounds prepared by ethylene glycol assisted sol-gel technique, viz. to itemize the structural phase transition from the polar rhombohedral to the anti-polar orthorhombic followed by the non-polar orthorhombic phase driven by the dopant content using both local scale and microscopic measurement techniques. The obtained results highlight the differences specific for the MPB assuming the data obtained by microscopic measurements as compared to local scale structural results.

## 2. Experimental

High purity precursors of Bi(NO$_3$)$_3$·5H$_2$O (99.99 %), Fe(NO$_3$)$_3$·9H$_2$O (99.95 %), Sm$_2$O$_3$ (99.99 %), 1,2-Ethanediol (99.8 %) and concentrated nitric acid were used for the synthesis. Firstly, an initial solution containing 50 mL of distilled water and 2 mL of the nitric acid was prepared and heated up to 80 °C. While under constant stirring the desired amount samarium oxide was dissolved, which was followed by the addition of the remaining metal precursor materials. The as obtained solution after all of the dissolution of the nitrates and Sm$_2$O$_3$ was left to stir for an additional hour. Lastly, ethylene glycol was added to the solution as to have a molar ratio of 2:1 as compared to the metal cations. The solution was then again left to stir for 1 hour after which all the liquid was evaporated by raising the temperature of the magnetic stirrer to 200 °C. The remaining gel was then dried in a drying furnace at 150 °C overnight and was then calcinated at 800 °C for 1.5 hours with a heating rate of 1 °C/min [20].

Rigaku MiniFlex diffractometer was used for the measurement of diffraction patterns. Which were recorded using Cu Kα radiation (λ=1.541874 Å) in 10º - 70º 2θ range, scanning speed was kept at 5 º/min with a step size of 0.02º. Hitachi SU-70 SEM was used for taking of scanning electron microscopy (SEM) micrographs . A dual beam system FE-SEM-FIB Helios Nanolab 650 with an energy dispersive X-ray (EDX) spectrometer INCA Energy 350 with X-Max 20 SDD detector was employed for measurement of chemical composition. Transmission electron microscopy (TEM) images were taken on FEI Tecnai G2 F20 X-TWIN. The XPS analysis was carried out with a Kratos Axis Supra spectrometer using monochromatic Al Kα source (25 mA, 15 kV). The instrument work function was calibrated to give a binding energy (BE) of 83.96 eV for the Au 4f7/2 line for metallic gold and the spectrometer dispersion was adjusted to give a BE of 932.62 eV for the Cu 2p3/2 line of metallic copper. Spectra were analyzed using CasaXPS software. Piezoresponse force microscopy technique was used to characterize local piezoelectric properties. Experiments have been carried out using MFP-3D commercial scanning probe microscope (Oxford Instruments, UK) using standard built-in mode with HA_HR Scansens commercial tips under *ac* voltage $V_{ac}$ = 3 V and frequency $f$ = 20 kHz. Both in-plane and out-of-plane PFM signals were measured. R*Cos(Θ) in-phase and R*Sin(Θ) out-of-phase piezoresponse signals were registered and treated by the subtracting of the frequency dependent background according to the procedure reported in Ref.[28].

## 3. Results and discussion

*3.1 XRD measurements*

XRD patterns recorded for the compounds $Bi_{1-x}Sm_xFeO_3$ have confirmed the formation of pure phase solid solutions without the presence of any impurity phases. The obtained results allowed to itemize the changes occurring in the crystal structure driven by an increase in the amount of the dopant ions. The XRD pattern obtained for the compound with samarium content of 8 mol % is consistent with a single-phase model with the rhombohedral structure (space group R3c) which is similar to the structure specific for the undoped compound $BiFeO_3$ [15]. An increase in the dopant concentration up to 12 mol % leads to a formation of $PbZrO_3$-type anti-polar orthorhombic structure (s.g. Pbam), however the content of the orthorhombic phase is relatively small (~ 8 %) as compared to the dominant rhombohedral phase formed in the compound (Fig. 1) [24].

Further increase in the dopant content leads to an increase in the intensity of the reflections attributed to the anti-polar orthorhombic structure, while the reflections attributed to the rhombohedral phase decrease confirming a reduction in the volume fraction of the R-phase down to ~ 20 % in the compound with x = 14 %. Larger amount of samarium content leads to a complete disappearance of the polar rhombohedral phase and a formation of new non-polar orthorhombic phase described by space group Pnma [19].

Thus the concentration range for the coexistence of the polar and the anti-polar phases is considered to be in the range $0.12 \leq x \leq 0.15$, however there are some previous reports where authors suggested single phase Pbam structure for the compounds having 15 mol % of samarium content [29, 30]. It should be noted, that the phase stability regions are notably dependent on the sample preparation techniques and other factors [31,32]. In the present study, the structural state of the compound containing 16 mol. % of the dopant ions is described by two-phase model considering a coexistence of two orthorhombic phases, viz. about 20 mol % of the non-polar phase described by space group Pnma and the dominant anti-polar orthorhombic phase (s.g. Pbam). Increase in the dopant content leads to a drastic increase in the volume fraction of the non-polar orthorhombic phase and the compound with x = 0.2 is considered to be single phase having pure non-polar orthorhombic structure.

Thus, in the mentioned concentration range there are two different phase coexistence regions, first region ($0.12 \leq x \leq 0.15$) is associated with a mixture of the polar and the anti-polar phases, the second region ($0.15 < x < 0.20$) is ascribed to a combination of two different orthorhombic phases (Pbam and Pnma). The phase coexistence regions are characterized by a modification in the grain morphology which minimizes the energy of the structural phase transitions [33, 34]. Crystallites visualized by SEM methods for the compounds within the phase boundary regions are characterized by reduced average size as compared to the crystallites in the compounds outside the MPB region [20] which is also confirmed by a broadening of the X-ray diffraction peaks recorded for these compounds [2, 30]. Thus, the XRD pattern of the compound with x = 0.08 is characterized by rather narrow diffraction peaks at around 39.5° and 45.5°. The compounds with larger amount of the dopant content attributed to the MPB regions are characterized by broadened diffraction peaks as seen by FWHM parameters at Figure 1. In this case of the compounds doped with 18 mol. % and 20 mol. % of samarium a narrowing of the reflections is observed which point at slight increase in the crystallite size. This tendency can be clearly seen from the SEM data (Figure 2 inset) and the results of XRD demonstrate that all of the marked peaks at 22.5°, 25.5° and 33.0° become narrower and sharper with a decreased FWHM.

*3.2 SEM measurements*

Grain size is the crucial issue for the compounds near the MPB as this factor drastically affects the structural stability of the compounds as well as their properties. The mentioned factor influences the stability of the intermediary phases as well the concentration range specific for the phase boundary regions [36,37].

Analysis of the SEM data allowed to evaluate the change in crystallite morphology as well as their average size as a function of the samarium content (Fig. 2). Particle size was estimated by ImageJ software [38] and is around 0.74 μm for the compound with 8 mol.% of samarium content. Further chemical doping leads to a significant decrease in the average grain size, it is around 0.52 μm for the compounds with x = 0.12 and 0.14. The smallest grain size of about 0.45 μm was observed for the compound with x = 0.16. The compounds specific for the end of the phase boundary region (18 mol. % and 20 mol. %) have similar grain size of around 0.50 μm. The observed change is the average grain size can be caused by several factors. One is associated with a fact that samarium ions show weaker chemical reactivity as compared to bismuth ions which in turn reduces the speed of mass transport and particle growth [39]. Other factor is associated with the fact that bismuth ions are characterized by high volatility which induces an increase in the amount of oxygen vacancies which in turn slow down grain growth [40]. It should be noted that the distribution of the individual particle size is rather broad (0.1 - 2.2 μm), while this is quite common for the compounds prepared by sol-gel method and heated at elevated temperatures [41,42] wherein the particle shape remain to be nearly similar (rectangular-like form) which corresponds to the geometry of the unit cell estimated for the compounds [43].

*3.3 EDS/EDX and XPS measurements*

In order to investigate the distribution of different chemical elements inside the crystallites of the compounds as well as depending on the dopant content the measurements using energy dispersive spectroscopy were performed (Figure S1). The obtained data indicate that there is no significant segregation of the ions across the crystallites, in particular near the grain boundaries as compared to some perovskite type compounds [44]. The distribution of chemical elements inside the grains is rather homogeneous for the compounds under study, viz. as for the compounds characterized by the phase coexistence state as for the single-phase compounds. The EDS results testify that the element content corresponds to the designed chemical formulas for the appropriate compounds, while a careful analysis of the EDS data has allowed to clarify a certain difference in the elements distribution as a function of the dopant content.

For the compound with 12 % of Sm content the difference in the Sm/Fe ratio is about 1.8 % (as compared to theoretical value) which confirms quite homogeneous distribution of the elements and corresponds to nearly single phase structural state of the compounds as confirmed by the XRD data. In the compound with x = 0.16 the mentioned difference is about 5.5 %, thus an average inaccuracy in the chemical composition is less than 1 % which still confirms high homogeneity in the chemical composition. The mentioned distribution in the chemical elements is in accordance with the two phase structural state estimated by the diffraction measurements. It should be noted that the mentioned difference in the elements distribution could not allow to reveal distinct crystallites ascribed to the different structural phases during SEM measurements which points at nanoscale character of the coexisting structural phases. The difference in the Sm/Fe ratio estimated for the compound with x = 0.2 is about 8.1 % which gives about 1.6 % of average inaccuracy in the chemical composition and provides variations in the crystal structure which are undetectable by the XRD data (Table S1). It should be noted that chemical homogeneity of the compounds is associated with optimal synthesis conditions, relatively low heating temperatures and short sintering time leading to low volatility of bismuth ions. It can also be stated that two phase structural state observed for the compounds within the phase boundary region is thermodynamically stable state and is not associated with a chemical inhomogeneity of the samples.

X-Ray photoelectron spectroscopy measurements performed for the compounds have confirmed the ratio of the chemical elements used during synthesis procedure while no additional elements were detected as confirmed from survey spectra (Figure S2). High resolution spectra were recorded to further investigate core levels of each principal element. Analysis of the Fe 2p spectra (Figure 3) has allowed to determine the oxidation state of iron ions. Iron 2p spectra showed characteristic broadening due to spin-orbital coupling between 2p core hole and unpaired 3d electrons of the photoionized Fe cation and also crystal field interactions [45]. All spectra show satellite peaks attributed to high-spin $Fe^{2+}$ and $Fe^{3+}$ oxidation states. Fe 2p splitting corresponding to high-spin $Fe^{3+}$ and $Fe^{2+}$ was modeled using Gupta and Sen multiplet predictions, high BE surface structures and satellite peaks for Fe 2p 3/2 region and are given in the Figure 3 [46]. Analysis of the XPS data has showed that Fe 2p spectra contained $Fe^{3+}$ peaks at 710.5 eV, 711.8 eV, 713.6 eV, 715.9 eV and $Fe^{2+}$ peaks at 709.5 eV, 710.4 eV, and 711.3 eV as well satellite peaks of $Fe^{2+}$ and $Fe^{3+}$. Calculated $Fe^{3+}$ and $Fe^{2+}$ ratio is 85 % and ~15 % and it remains nearly stable for the compounds with x < 0.12. Increase in the dopant content leads to an increase in the amount of $Fe^{2+}$ ions up to about 22 %.

Analysis performed for the 4f spectra of Bi ions has indicated that each of Bi 4f peaks were composed of two separate components those ratio remains similar for doping up to x = 0.14. Increase in the dopant concentration leads to notable increase of lower energy component area. This significant change possibly corresponds to the symmetry change in the compounds as well and the decrease of oxygen vacancy content observed in O1s spectra. Oxygen spectra also show an asymmetric peak shape, which testify a formation of oxygen defects in the compounds with x > 0.14 as seen on the Figures S3, S4 and S5 [47]. Even though the existence of large amount of $Fe^{2+}$ ions and oxygen vacancies were measured to exist, Rietveld refinement results did not coincide with such quantities. And XPS results were used more as an indication for the detection of Sm influence rather than absolute value determination.

*3.4 TEM measurements*

In order to specify the concentration driven evolution of the crystal structure of the compounds with chemical compositions across the morphotropic phase boundary region the HR-TEM measurements have been performed at room temperature. The HR-TEM measurements have been carried out for the compounds with x = 0.08, 0.12, 0.16 and 0.20 which allow to trace the concentration driven structural transition from the rhombohedral phase to the single phase non-polar orthorhombic one via two phase regions ascribed to the polar - anti-polar and anti-polar – non-polar phase boundaries. The TEM images correspond to single crystallites which compose of different structural phases (Figure 4).

The FFT (Fast Fourier Transforms) performed for the compound with x = 0.08 testify the rhombohedral distortion of the crystal structure estimated by the diffraction measurements. While along with the dominant rhombohedral phase, some areas of the crystallites contain the inclusions which correspond to the orthorhombic phase (Fig. 4, inset). The obtained TEM data could not allow to specify a type of the orthorhombic structure while assuming the results of the diffraction measurements we consider the anti-polar orthorhombic described by Pbam space group as most probable secondary phase. The TEM data obtained for the compound with x = 0.12 testify more often presence of the orthorhombic phase which can be observed in one crystallite (Fig. 4, inset) thus confirming nanoscale size of the phase coexistence phenomena in the compounds near the morphotropic phase boundary. The TEM data obtained for the compound with x = 0.16 have allowed to specify the symmetry of the orthorhombic inclusions present in the crystallites. Analysis of the FFT images has allowed to itemize both the anti-polar and the non-polar orthorhombic symmetry thus confirming the results of XRD measurements refined using the space groups Pbam and Pnma respectively. The TEM data obtained for the compound with x = 0.2 confirm the dominance of the non-polar orthorhombic phase while FFT analysis performed for a number of crystallites has allowed to find out the areas of particular crystallites containing the rhombohedral phase specific for the lightly-doped compounds. Thus the TEM measurements have allowed to study in detail the structural phase transitions across the morphotropic phase boundary region and to itemize the changes occurred in the particular structural state on local scale level. The obtained results highlight the difference in the concentration ranges attributed to the different structural phases as compared to the results obtained by the XRD measurements.

*3.5 PFM measurements*

The results of the local piezoelectric measurements performed by PFM method demonstrate an evolution of the distribution and level of the piezoresponse as a function of the dopant concentration in the $Bi_{1-x}Sm_xFeO_3$ compounds. The average grain size of the compounds characterized by a dominance of the rhombohedral phase is ~1 μm which is much larger than an average size of polar domains. Increase in the dopant content leads to a notable reduction in the average domain size as well as a decrease of the crystallites size which confirms the results of the SEM and XRD measurements. It was found that out-of-plane PFM signal contains large unipolar contribution and thereby further analysis was performed on the in-plane PFM images, where unipolar electromechanical contribution is much lower and can be removed by the mathematical procedure discussed in Ref. [28]. Polar active regions can be identified on the PFM images (Figure 5) as areas with bright and dark contrasts with the amplitude of the piezoresponse signal larger than 5 pm. The change of the piezoresponse sign reveals differently distributed polarization inside the crystallites. It should be noted that the PFM contrast images obtained for the compound with x = 0.08 are characterized by a presence of the regions with close-to-zero PFM signal (below 2 pm) showing the presence of the non-polar or the anti-polar phase inclusions. Taking into account the XRD and TEM data one can conclude about the presence of the nanoscale fraction of the anti-polar orthorhombic phase in the compound with x = 0.08 while the compound is characterized by single phase structural state assuming solely the X-ray diffraction data.

PFM results demonstrate that an increase in the dopant content leads to an increase in the area of polar neutral regions which correspond to an increase in the anti-polar and the non-polar orthorhombic phases as confirmed by the XRD data. Increase in the samarium content also leads to a reduction in the value of piezoresponse signal wherein the domains reduce in size and become to be of curved shape with smeared borders. A degradation of the PFM signal is in accordance with the XRD results which confirm a decrease in the amount of the polar rhombohedral phase along with an increase in the anti-polar phase followed by the stabilization of the non-polar orthorhombic single phase state. The polar active regions revealed by the PFM measurements in the compounds characterized by a coexistence of the orthorhombic phases (viz. with x = 0.16, 0.18) possess nanoscale size domains. The PFM data obtained for the single phase non-polar orthorhombic compound with x = 0.20 also testify a presence of polar active regions with small piezoresponse signal. The distribution of the signal also reveals a decrease in the polar phase area following by a "diffusion" of the PFM images and "erasing" the boundaries between the polar and non-polar areas. The presence of the polar active regions in the compounds with no rhombohedral phase as confirmed by the XRD data can be explained assuming the following two scenarios. The first one assumes an induction of piezoelectric signal during the measurements due to a small difference in the ground state energies ascribed to the anti-polar and the polar active phases [2], the value of the mentioned energy barrier is comparable with an electric field intensity formed within the samples surface region ($\Delta E \sim 0.5$ MV/cm) during PFM measurements [48]. The ability of the $BiFeO_3$-based materials to exhibit electric field induced was earlier confirmed both in microscopic XRD [2] and local PFM measurements in solid phase sintered ceramics [44]. The second scenario assumes a presence of small amount of polar active phase which volume fraction is less than 2-3 % percent, the mentioned polar active phase has mainly 2D character with random distribution over the crystallites surface which have dominant orthorhombic state thus remaining to be undetectable by conventional XRD methods. It should be noted that the results of TEM measurements make the latter scenario to be more preferable which considers a presence of the rhombohedral and orthorhombic inclusions in the compounds with dominant orthorhombic and rhombohedral phases respectively.

## Conclusions

Sm-substituted $Bi_{1-x}Sm_xFeO_3$ ceramics prepared by a modified sol-gel method are characterized by nanoscale size crystallites which average size slightly decreases with the increase of dopant content. Increase in samarium concentration leads to the structural transition from the polar active rhombohedral phase to the non-polar orthorhombic phase via a formation of metastable anti-polar orthorhombic phase. The morphotropic phase boundary region determined from the diffraction results is characterized by two concentration ranges having coexistent polar rhombohedral and anti-polar orthorhombic phases ($0.12 \leq x \leq 0.15$) and the anti-polar and non-polar orthorhombic phases ($0.15 < x < 0.18$). Local scale measurements confirmed that morphotropic phase boundary region is characterized by wider concentration range, viz. $0.08 \leq x \leq 0.20$. Notable difference with the results of microscopic measurements is justified by the presence of the polar active and the non-polar regions in the compounds having single phase non-polar orthorhombic ($x > 0.18$) and the polar rhombohedral structural state ($x < 0.12$) respectively as determined by the diffraction measurements. The mentioned structural regions are characterized by a volume fraction of about 2 - 3 % which coexist along the dominant structural phase and randomly distributed over a surface of crystallites ascribed to the major phases.

## Acknowledgements

This work was supported by the European Union's Horizon 2020 research and innovation programme under the Marie Skłodowska-Curie grant agreement No. 778070. M.V.S acknowledges Ministry of Science and Higher Education of the Russian Federation within the framework of state support for the creation and development of World-Class Research Centers "Digital biodesign and personalized healthcare" №075-15-2020-926. Diffraction measurements and analysis (A.A.D. and D.V.K.) were supported by RFBR (projects # 20-58-00030) and BRFFR (project # F20R-123). Piezoresponse force microscopy investigations were made possible by the Russian Science Foundation (grant 19-72-10076). The equipment of the Ural Center for Shared Use "Modern nanotechnology" UrFU was used.

## References

[1]  M. Ahart, M. Somayazulu, R.E. Cohen, P. Ganesh, P. Dera, H. Mao, R.J. Hemley, Y. Ren, P. Liermann, Z. Wu, Origin of morphotropic phase boundaries in ferroelectrics., Nature. 451 (2008) 545–8. https://doi.org/10.1038/nature06459.

[2]  J. Walker, H. Simons, D.O. Alikin, A.P. Turygin, V.Y. Shur, A.L. Kholkin, H. Ursic, A. Bencan, B. Malic, V. Nagarajan, T. Rojac, Dual strain mechanisms in a lead-free morphotropic phase boundary ferroelectric, Sci. Rep. 6 (2016) 1–8. https://doi.org/10.1038/srep19630.

[3]  F. Zheng, J. Chen, X. Li, M. Shen, Morphotropic phase boundary (MPB) effect in Pb (Zr,Ti)O3 rhombohedral/tetragonal multilayered films, Mater. Lett. 60 (2006) 2733–2737. https://doi.org/10.1016/j.matlet.2006.01.080.

[4]  H. Zheng, I.M. Reaney, W.E. Lee, N. Jones, H. Thomas, Effects of strontium substitution in Nb-doped PZT ceramics, J. Eur. Ceram. Soc. 21 (2001) 1371–1375. https://doi.org/10.1016/S0955-2219(01)00021-8.

[5]  B. Noheda, D.E. Cox, G. Shirane, J.A. Gonzalo, L.E. Cross, S.E. Park, A monoclinic ferroelectric phase in the Pb(Zr1_xTix)O 3 solid solution, Appl. Phys. Lett. 74 (1999) 2059–2061. https://doi.org/10.1063/1.123756.

[6]  D. Damjanovic, Comments on origins of enhanced piezoelectric properties in ferroelectrics, in: IEEE Trans. Ultrason. Ferroelectr. Freq. Control, 2009: pp. 1574–1585. https://doi.org/10.1109/TUFFC.2009.1222.

[7]  D. Damjanovic, Lead-based piezoelectric materials, in: Piezoelectric Acoust. Mater. Transducer Appl., Springer US, 2008: pp. 59–79. https://doi.org/10.1007/978-0-387-76540-2_4.

[8]  J. Rödel, J.F. Li, Lead-free piezoceramics: Status and perspectives, MRS Bull. 43 (2018) 576–580. https://doi.org/10.1557/mrs.2018.181.

[9] J. Rödel, W. Jo, K.T.P. Seifert, E.M. Anton, T. Granzow, D. Damjanovic, Perspective on the development of lead-free piezoceramics, J. Am. Ceram. Soc. 92 (2009) 1153–1177. https://doi.org/10.1111/j.1551-2916.2009.03061.x.

[10] D. Damjanovic, A morphotropic phase boundary system based on polarization rotation and polarization extension, Appl. Phys. Lett. 97 (2010). https://doi.org/10.1063/1.3479479.

[11] M.R. Suchomel, P.K. Davies, Predicting the position of the morphotropic phase boundary in high temperature $PbTiO_3$-$Bi(B'B''')O_3$ based dielectric ceramics, J. Appl. Phys. 96 (2004) 4405–4410. https://doi.org/10.1063/1.1789267.

[12] T. Zheng, J. Wu, D. Xiao, J. Zhu, Recent development in lead-free perovskite piezoelectric bulk materials, Prog. Mater. Sci. 98 (2018) 552–624. https://doi.org/10.1016/j.pmatsci.2018.06.002.

[13] C.M. Fernández-Posada, A. Castro, J.-M. Kiat, F. Porcher, O. Peña, R. Jiménez, M. Algueró, H. Amorín, The Polar/Antipolar Phase Boundary of $BiMnO_3$-$BiFeO_3$-$PbTiO_3$: Interplay among Crystal Structure, Point Defects, and Multiferroism, Adv. Funct. Mater. 28 (2018) 1802338. https://doi.org/10.1002/adfm.201802338.

[14] D. V. Karpinsky, I.O. Troyanchuk, M. Tovar, V. Sikolenko, V. Efimov, A.L. Kholkin, Evolution of crystal structure and ferroic properties of La-doped BiFeO3 ceramics near the rhombohedral-orthorhombic phase boundary, J. Alloys Compd. 555 (2013) 101–107. https://doi.org/10.1016/j.jallcom.2012.12.055.

[15] T. Rojac, A. Bencan, B. Malic, G. Tutuncu, J.L. Jones, J.E. Daniels, D. Damjanovic, $BiFeO_3$ Ceramics: Processing, Electrical, and Electromechanical Properties, J. Am. Ceram. Soc. 97 (2014) 1993–2011. https://doi.org/10.1111/jace.12982.

[16] S.H. Han, K.S. Kim, H.G. Kim, H.G. Lee, H.W. Kang, J.S. Kim, C. Il Cheon, Synthesis and characterization of multiferroic BiFeO3 powders fabricated by hydrothermal method, Ceram. Int. 36 (2010) 1365–1372. https://doi.org/10.1016/j.ceramint.2010.01.020.

[17] D. Lebeugle, D. Colson, A. Forget, M. Viret, A.M. Bataille, A. Gukasov, Electric-field-induced spin flop in BiFeO3 single crystals at room temperature, Phys. Rev. Lett. 100 (2008) 227602. https://doi.org/10.1103/PhysRevLett.100.227602.

[18] W. Xing, M. Yinina, Z. Ma, Y. Bai, J. Chen, S. Zhao, Improved ferroelectric and leakage current properties of Er-doped BiFeO3 thin films derived from structural transformation, Smart Mater. Struct. 23 (2014) 85030. https://doi.org/10.1088/0964-1726/23/8/085030.

[19] I.O. Troyanchuk, D. V. Karpinsky, M. V. Bushinsky, O.S. Mantytskaya, N. V. Tereshko, V.N. Shut, Phase transitions, magnetic and piezoelectric properties of rare-earth-substituted BiFeO3ceramics, J. Am. Ceram. Soc. 94 (2011) 4502–4506. https://doi.org/10.1111/j.1551-2916.2011.04780.x.

[20] D. V. Karpinsky, A. Pakalniškis, G. Niaura, D. V. Zhaludkevich, A.L. Zhaludkevich, S.I. Latushka, M. Silibin, M. Serdechnova, V.M. Garamus, A. Lukowiak, W. Stręk, M. Kaya, R. Skaudžius, A. Kareiva, Evolution of the crystal structure and magnetic properties of Sm-doped BiFeO3 ceramics across the phase boundary region, Ceram. Int. (2020). https://doi.org/10.1016/j.ceramint.2020.10.120.

[21] M. Kubota, K. Oka, Y. Nakamura, H. Yabuta, K. Miura, Y. Shimakawa, M. Azuma, Sequential Phase Transitions in Sm Substituted $BiFeO_3$, Jpn. J. Appl. Phys. 50 (2011) 09NE08. https://doi.org/10.7567/jjap.50.09ne08.

[22] A. Herklotz, S.F. Rus, N. Balke, C. Rouleau, E.J. Guo, A. Huon, S. Kc, R. Roth, X. Yang, C. Vaswani, J. Wang, P.P. Orth, M.S. Scheurer, T.Z. Ward, Designing Morphotropic Phase Composition in BiFeO3, Nano Lett. 19 (2019) 1033–1038. https://doi.org/10.1021/acs.nanolett.8b04322.

[23] S. Gupta, S. Bhattacharjee, D. Pandey, V. Bansal, S.K. Bhargava, J.L. Peng, A. Garg, Absence of morphotropic phase boundary effects in BiFeO3- PbTiO3 thin films grown via a chemical multilayer deposition method, Appl. Phys. A Mater. Sci. Process. 104 (2011) 395–400. https://doi.org/10.1007/s00339-010-6163-5.

[24] D. Arnold, Composition-driven structural phase transitions in rare-earth-doped bifeo3 ceramics: A review, IEEE Trans. Ultrason. Ferroelectr. Freq. Control. 62 (2015) 62–82. https://doi.org/10.1109/TUFFC.2014.006668.

[25] S. Bhattacharyya, J.R. Jinschek, H. Cao, Y.U. Wang, J. Li, D. Viehland, Direct high-resolution transmission electron microscopy observation of tetragonal nanotwins within the monoclinic MC phase of Pb (Mg13 Nb23) O3 -0.35PbTi O3 crystals, Appl. Phys. Lett. 92 (2008) 142904. https://doi.org/10.1063/1.2908228.

[26] K.A. Schönau, L.A. Schmitt, M. Knapp, H. Fuess, R.A. Eichel, H. Kungl, M.J. Hoffmann, Nanodomain structure of Pb [Zr1-x Tix] O3 at its morphotropic phase boundary: Investigations from local to average structure, Phys. Rev. B - Condens. Matter Mater. Phys. 75 (2007) 184117. https://doi.org/10.1103/PhysRevB.75.184117.

[27] A. Pakalniškis, A. Lukowiak, G. Niaura, P. Głuchowski, D.V. Karpinsky, D.O. Alikin, A.S. Abramov, A. Zhaludkevich, M. Silibin, A.L. Kholkin, R. Skaudžius, W. Strek, A. Kareiva, Nanoscale ferroelectricity in pseudo-cubic sol-gel derived barium titanate - bismuth ferrite (BaTiO3– BiFeO3) solid solutions, J. Alloys Compd. (2020) 154632. https://doi.org/10.1016/j.jallcom.2020.154632.

[28] T. JUNGK, Á. HOFFMANN, E. SOERGEL, Consequences of the background in piezoresponse force microscopy on the imaging of ferroelectric domain structures, J. Microsc. 227 (2007) 72–78. https://doi.org/10.1111/j.1365-2818.2007.01783.x.

[29] V.A. Khomchenko, J.A. Paixão, B.F.O. Costa, D. V. Karpinsky, A.L. Kholkin, I.O. Troyanchuk, V. V. Shvartsman, P. Borisov, W. Kleemann, Structural, ferroelectric and magnetic properties of Bi0.85Sm0.15FeO3 perovskite, Cryst. Res. Technol. 46 (2011) 238–242. https://doi.org/10.1002/crat.201100040.

[30] S. Pattanayak, R.N.P. Choudhary, P.R. Das, Effect of Sm-substitution on structural, electrical and magnetic properties of BiFeO3, Electron. Mater. Lett. 10 (2014) 165–172. https://doi.org/10.1007/s13391-013-3050-1.

[31] X. Bai, Size and doping effect on the structure, transitions and optical properties of multiferroic BiFeO₃ particles for photocatalytic applications, 2016.

[32] Y. Zheng, R. Ran, Z. Shao, Cr doping effect in B-site of La0.75Sr0.25MnO 3 on its phase stability and performance as an SOFC anode, Rare Met. 28 (2009) 361–366. https://doi.org/10.1007/s12598-009-0072-9.

[33] B. Molleman, T. Hiemstra, Size and shape dependency of the surface energy of metallic nanoparticles: Unifying the atomic and thermodynamic approaches, Phys. Chem. Chem. Phys. 20 (2018) 20575–20587. https://doi.org/10.1039/c8cp02346h.

[34] J. Cahn, J.W. Cahn, Transitions and phase equilibria among grain boundary structures, J. Phys. Colloq. 43 (1982). https://doi.org/10.1051/jphyscol:1982619ï.

[35] R. Sumang, N. Vittayakorn, T. Bongkarn, Crystal structure, microstructure and electrical properties of (1-x-y)Bi0.5Na0.5TiO3-xBi0.5K 0.5TiO3-yBiFeO3 ceramics near MPB prepared via the combustion technique, in: Ceram. Int., Elsevier, 2013: pp. S409–S413. https://doi.org/10.1016/j.ceramint.2012.10.104.

[36] C. Behera, R.N.P. Choudhary, P.R. Das, Structural, electrical and multiferroic characteristics of thermo-mechanically fabricated BiFeO3-(BaSr)TiO3 solid solutions, Mater. Res. Express. 5 (2018) 056301. https://doi.org/10.1088/2053-1591/aabeae.

[37] W. Jo, T.H. Kim, D.Y. Kim, S.K. Pabi, Effects of grain size on the dielectric properties of Pb (Mg1/3 Nb2/3) O3 -30 mol % PbTiO3 ceramics, J. Appl. Phys. 102 (2007) 074116. https://doi.org/10.1063/1.2794377.

[38] J. Schindelin, I. Arganda-Carreras, E. Frise, V. Kaynig, M. Longair, T. Pietzsch, S. Preibisch, C. Rueden, S. Saalfeld, B. Schmid, J.Y. Tinevez, D.J. White, V. Hartenstein, K. Eliceiri, P. Tomancak, A. Cardona, Fiji: An open-source platform for biological-image analysis, Nat. Methods. 9 (2012) 676–682. https://doi.org/10.1038/nmeth.2019.

[39] S. Cho, C. Yun, Y.S. Kim, H. Wang, J. Jian, W. Zhang, J. Huang, X. Wang, H. Wang, J.L. MacManus-Driscoll, Strongly enhanced dielectric and energy storage properties in lead-free perovskite titanate thin films by alloying, Nano Energy. 45 (2018) 398–406. https://doi.org/10.1016/j.nanoen.2018.01.003.

[40] S. Liu, H. Luo, S. Yan, L. Yao, J. He, Y. Li, L. He, S. Huang, L. Deng, Effect of Nd-doping on structure and microwave electromagnetic properties of BiFeO3, J. Magn. Magn. Mater. 426 (2017) 267–272. https://doi.org/10.1016/j.jmmm.2016.11.080.

[41] L. V. Yafarova, I. V. Chislova, I.A. Zvereva, T.A. Kryuchkova, V. V. Kost, T.F. Sheshko, Sol–gel synthesis and investigation of catalysts on the basis of perovskite-type oxides GdMO3 (M = Fe, Co), J. Sol-Gel Sci. Technol. 92 (2019) 264–272. https://doi.org/10.1007/s10971-019-05013-3.

[42] M. Catauro, E. Tranquillo, G. Dal Poggetto, M. Pasquali, A. Dell'Era, S.V. Ciprioti, Influence of the heat treatment on the particles size and on the crystalline phase of TiO2 synthesized by the sol-gel method, Materials (Basel). 11 (2018). https://doi.org/10.3390/ma11122364.

[43] M.K. Singh, First principle study of crystal growth morphology: An application to crystalline urea, (2006). http://arxiv.org/abs/cond-mat/0602385 (accessed July 29, 2020).

[44] D.O. Alikin, A.P. Turygin, J. Walker, A. Bencan, B. Malic, T. Rojac, V.Y. Shur, A.L. Kholkin, The effect of phase assemblages, grain boundaries and domain structure on the local switching behavior of rare-earth modified bismuth ferrite ceramics, Acta Mater. 125 (2017) 265–273. https://doi.org/10.1016/j.actamat.2016.11.063.

[45] A.P. Grosvenor, B.A. Kobe, M.C. Biesinger, N.S. McIntyre, Investigation of multiplet splitting of Fe 2p XPS spectra and bonding in iron compounds, Surf. Interface Anal. 36 (2004) 1564–1574. https://doi.org/10.1002/sia.1984.

[46] R.P. Gupta, S.K. Sen, Calculation of multiplet structure of core p -vacancy levels. II, Phys. Rev. B. 12 (1975) 15–19. https://doi.org/10.1103/PhysRevB.12.15.

[47] Y. Li, M.S. Cao, D.W. Wang, J. Yuan, High-efficiency and dynamic stable electromagnetic wave attenuation for la doped bismuth ferrite at elevated temperature and gigahertz frequency, RSC Adv. 5 (2015) 77184–77191. https://doi.org/10.1039/c5ra15458h.

[48] D. Kan, L. Pálová, V. Anbusathaiah, C.J. Cheng, S. Fujino, V. Nagarajan, K.M. Rabe, I. Takeuchi, Universal behavior and electric-field-lnduced structural transition in rare-earth-substituted BiFeO3, Adv. Funct. Mater. 20 (2010) 1108–1115. https://doi.org/10.1002/adfm.200902017.

# Figures

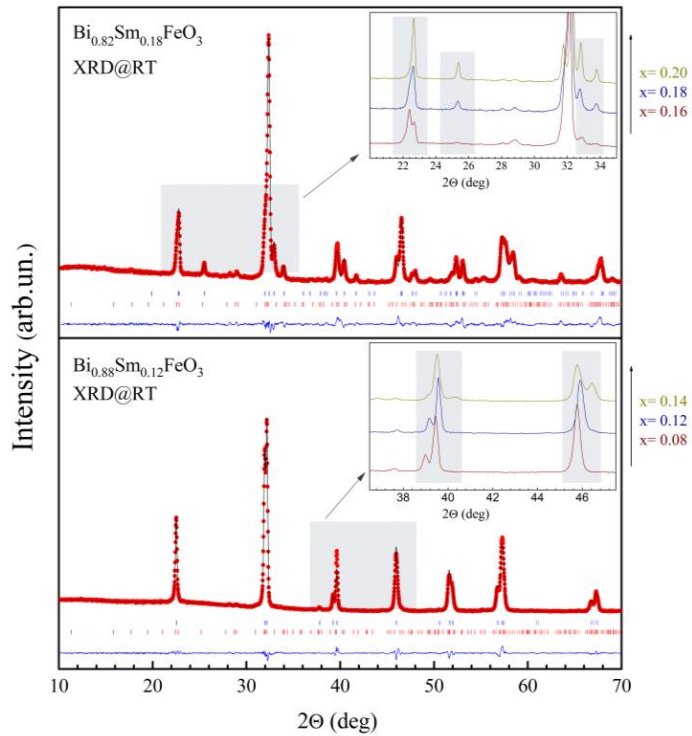

Fig. 1. XRD patterns of Bi1-xSmxFeO3 compounds with x = 0.12 and x = 0.18 refined using the two phase models (upper Bragg ticks denote R3c and Pbam phases respectively for compounds with x=0.12 and 0.18; bottom row denotes Pnma phase). The insets show an evolution of the selected diffraction peaks for compounds under study.

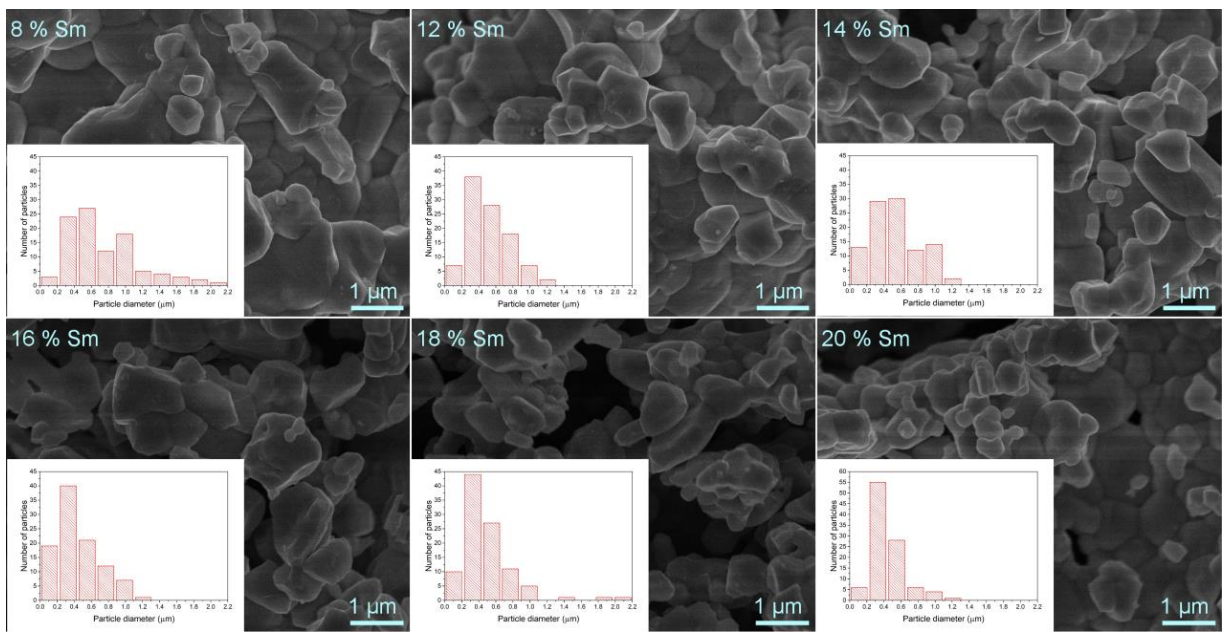

Fig. 2. SEM images of Bi1-xSmxFeO3 compounds x = 0.08, 0.12, 0.14, 0.16, 0.18, 0.20. The insets show particle size histograms for the compounds under study.

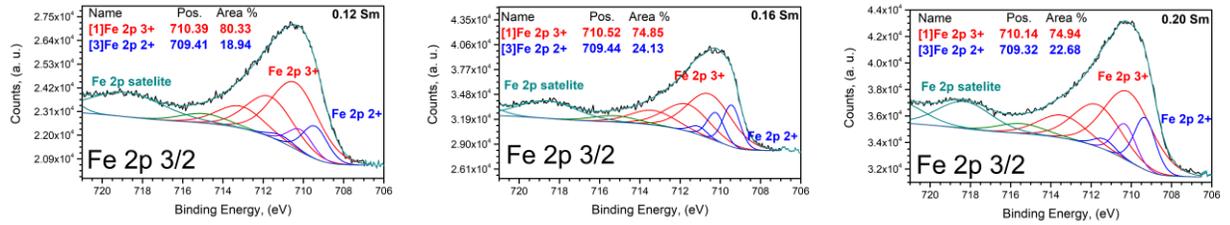

Fig. 3. Fitted high resolution XPS Fe 2p orbital spectra of $Bi_{1-x}Sm_xFeO_3$ compounds with x = 0.12, 0.16, 0.20 in the binding energy range from 538 to 524 eV.

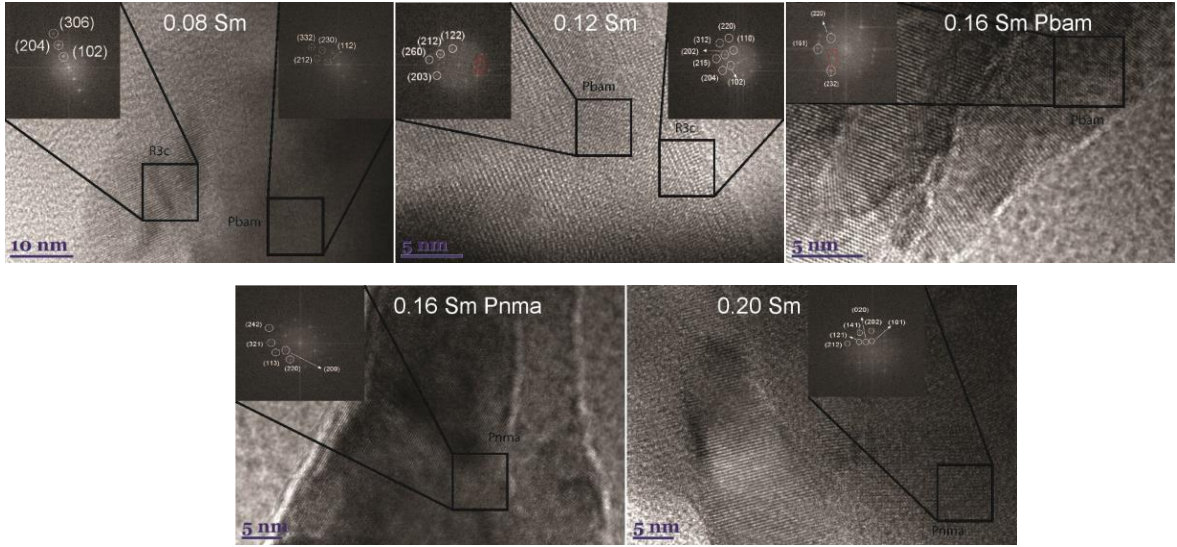

Fig. 4. HRTEM image of the compounds Bi1-xSmxFeO3 with x = 0.08, 0.12 and 0.16 0.20 (from left to right) at room temperature denoting a coexistence of different strucutral phases. The insents show FFT results calculated for different area of TEM images which testify a coexistence of different structural phases. Dots specific for Pbam phase caused by unit cell multiplication are ringed in red.

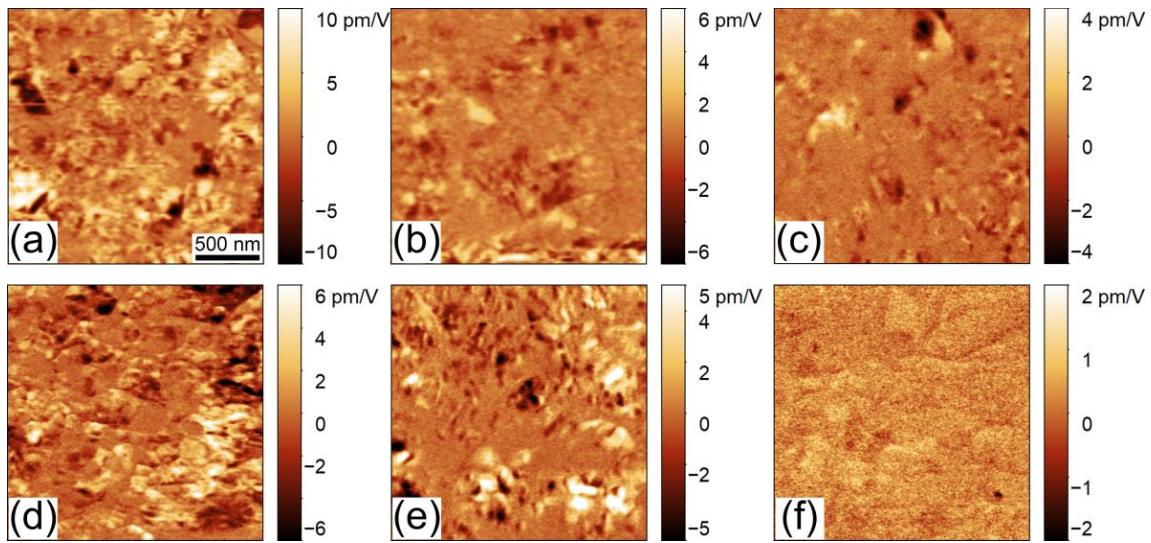

Fig. 5. In-plane PFM images of Bi1-xSmxFeO3 compounds with x = 8%, 12%, 14%, 16%, 18, 20 respectively for a) – f) pictures (from left to right and from up to down).